\documentclass{PoS}
\usepackage{epsfig}
\usepackage{array}
\def\dt{\delta\tau}                 
\def\dH{\delta H}                   
\def\trjlen{\tau}                   
\def\defn{\equiv}                   
\def\O{{\cal O}}                    
\def\F{\mathcal F}		    
\def\G{G}			    
\def\M{\mathcal M}		    
\def\hats{\hat S}                   
\def\hatt{\hat T}                   
\def\ST{\{S,T\}}                    
\def\SST{\{S,\ST\}}                 
\def\TST{\{T,\ST\}}                 
\def\TTTST{\{T,\{T,\{T,\ST\}\}\}}   
\def\STSST{\{\ST,\{S,\ST\}\}}       
\def\TSSST{\{T,\{S,\{S,\ST\}\}\}}   
\def\TTSST{\{T,\{T,\{S,\ST\}\}\}}   
\def\SSSST{\{S,\{S,\{S,\ST\}\}\}}   
\def\STTST{\{\ST,\{T,\ST\}\}}       
\def\rational#1#2{{\mathchoice{\textstyle{#1\over#2}}%
  {\scriptstyle{#1\over#2}}{\scriptscriptstyle{#1\over#2}}{#1/#2}}}
\def\half{\rational12}		
\title{Force Gradient Integrators}
\ShortTitle{Force Gradient Integrators}
\author{\speaker{A.~D.~Kennedy}\\
        School of Physics \& Astronomy\\
	University of Edinburgh\\
        E-mail: \email{adk@ph.ed.ac.uk}}
\author{M.~A.~Clark\\
        Harvard--Smithsonian Center for Astrophysics\\
	and\\
	Initiative in Innovative Computing\\
	Harvard University School of Engineering and Applied Sciences\\
	E-mail: \email{mikec@seas.harvard.edu}}
\author{P.~J.~Silva\\
        School of Physics \& Astronomy\\
	University of Edinburgh\\
        E-mail: \email{psilva@ph.ed.ac.uk}}

\abstract{We present initial results of the use of Force Gradient integrators
  for lattice field theories.  These promise to give significant performance
  improvements, especially for light fermions and large lattices.  Our results
  show that this is indeed the case, indicating a speed-up of more than a
  factor of two, which is expected to increase as the integration step size
  becomes smaller for larger lattices and smaller fermion masses.}

\FullConference{The XXVII International Symposium on Lattice Field
		 Theory - LAT2009\\
		 July 26-31 2009\\
		 Peking University, Beijing, China}
\begin{document}

\section{Introduction}

Our goal is to construct more accurate molecular dynamics integrators for use
in Hybrid Monte Carlo (HMC) computations for lattice quantum field theories.
In particular we shall consider symmetric symplectic integrators whose errors
are of higher order in the integration step size than those of the leapfrog
(also known as the St\"ormer or Verlet) method.  These integrators improve the
scaling behaviour \(\dH=k \cdot\dt^n\) from \(n=2\) to \(n=4\), which reduces
the cost for large enough volume and small enough fermion masses for which
\(\dt\to0\).  Previous methods for constructing higher-order integrators
\cite{campostrini89a,creutz89a} were thwarted by large values for the
coefficient \(k\); the new idea considered here is to compute second
derivatives ``analytically'' rather than ``numerically''
\cite{chin:2000,omelyan:2002}.

\section{Symplectic Integrators}

As described in \cite{Kennedy:2007cb,Clark:2008gh,Clark:2007ff} we may define
a Hamiltonian system for a gauge field by introducing the symplectic
fundamental \(2\)-form \(\omega\defn -d(p^i\theta_i)\) with \(\theta_i\) being
the frame of left-invariant Maurer-Cartan forms; this ensures that the
Hamiltonian dynamics is gauge invariant.  For every \(0\)-form \(F\) on phase
space this defines a Hamiltonian vector field \(\hat F\) satisfying
\(dF=i_{\hat F}\omega\), and the Hamiltonian evolution for the system
corresponds to an integral curve of the Hamiltonian vector field \(\hat H\)
for the Hamiltonian function \(H\).  We can find a closed-form integral curve
of \(\hat F\), that is evaluate \(e^{\hat F\trjlen}\) explicitly, if \(F\)
depends only on the ``positions'' (fields) \(q\) or momenta~\(p\).  This is
particularly useful when the Hamiltonian is of the form \(H(q,p)=S(q)+T(p)\)
as then we can integrate the Hamiltonian vector fields \(\hats\) and \(\hatt\)
exactly.

\section{Shadow Hamiltonians and Force Gradient Integrators}

We recall the Baker--Campbell--Hausdorff (BCH) formula, which states that if
\(A\) and \(B\) belong to any (in general non-commutative) associative algebra
then \(e^Ae^B = e^{A+B+\delta}\) where \(\delta\) is in the free Lie algebra
generated by \(A\) and \(B\).  Furthermore it gives an explicit expansion for
\(\delta\), and correspondingly for the symmetric product \[\ln\left(e^{A/2}
e^{B}e^{A/2} \right) = A + B - \frac1{24} \Bigl(\left[A,[A,B]\right] +
2\left[B,[A,B]\right]\Bigr) +\cdots.\]

Using the Jacobi identity one may show that the commutator of two Hamiltonian
vector fields is itself a Hamiltonian vector field, \([\hats,\hatt] =
\widehat{\ST}\), where \(\ST = -\omega(\hats,\hatt)\) is the Poisson bracket
of the two \(0\)-forms \(S\) and~\(T\).  We therefore find that any integrator
constructed from a sequence of symplectic steps exactly conserves a shadow
Hamiltonian \(\tilde H\) obtained from the BCH formula by replacing
commutators with Poisson brackets.

As a very simple example consider the PQPQP integrator \[\left(
e^{\alpha\hats\dt} e^{\half\hatt\dt} e^{(1-2\alpha)\hats\dt} e^{\half\hatt\dt}
e^{\alpha\hats\dt} \right)^{\trjlen/\dt}\] whose shadow Hamiltonian is
\[\tilde H = H + \left(\frac{6\alpha^2 - 6\alpha + 1}{12} \SST + \frac{1 -
6\alpha}{24} \TST \right)\dt^2 + \O(\dt^4).\]

As the Poisson bracket \(\SST\) does not depend on momentum we can integrate
the Hamiltonian vector field \(\widehat{\SST}\) exactly, and this ``second
derivative'' corresponds to the \emph{Force Gradient} just as the Hamiltonian
vector field \(\hats\) corresponds to the ``force''.  The explicit form of the
shadow Hamiltonian for a variety of integrators is shown in
Table~\ref{table:1}, the simplest Force Gradient integrator is given in the
last entry.

\begin{table}
  \begin{center}
    \setbox0=\hbox{} \ht0=0pt \dp0=0pt
    \newif\ifexp \expfalse
    \def\1#1#2{\ifexp\exp\left(#1\dt\,\hat#2\right)\else e^{#1\dt\,\hat#2}\fi}
    \def\2#1#2{\(#1\;#2\;#1\)}
    \def\3#1#2#3{\(\begin{array}{c}
          #1\;#2 \\
          \times\;#3 \\
	  \times\;#2\;#1
	\end{array}\)}
    \def\4#1#2#3#4{\(\setlength{\extrarowheight}{2ex}\begin{array}{c}
	  #1\\
	  \times\;#2\\
	  \times\;#3 \\
          \times\;#4 \\
	  \times\;#3 \\
	  \times\;#2 \\
	  \times\;#1
	\end{array}\)}
    \def\7#1#2#3#4{\(\setlength{\extrarowheight}{2ex}\begin{array}{c}
	  #1\;#2\;#3 \\
          \times\;#4 \\
	  \times\;#3\;#2\;#1
	\end{array}\)}
    \def\5#1#2#3{\mathchoice{{\textstyle\6{#1}{#2}{#3}}}%
      {{\scriptstyle\6{#1}{#2}{#3}}}%
      {{\scriptscriptstyle\6{#1}{#2}{#3}}}%
      {{\scriptscriptstyle\6{#1}{#2}{#3}}}}
    \newif\ifprev
    \newif\ifunit
    \def\8#1#2{
      \ifcat x#2x\unittrue\else\unitfalse\fi
      \count0=#1
      \ifnum\count0=0\else
        \ifnum\count0>0 \ifprev+\fi
	  \ifunit\the\count0\else\ifnum\count0>1 \the\count0\fi\fi
	\fi
	\count0=-\count0
        \ifnum\count0>0 -
	  \ifunit\the\count0\else\ifnum\count0>1 \the\count0\fi\fi
	\fi
	\ifunit\else#2\fi\prevtrue
      \fi}
    \def\6#1#2#3{\8{#1}{\root3\of4}\8{#2}{\root3\of2}\8{#3}{}}
    \begin{tabular}{c|c|c}
      Integrator & Update steps & Shadow Hamiltonian \\
      \noalign{\hrule height1.5pt}
      PQP & \2{\1\half S}{\1{}T}
        & \(T + S - \frac{\dt^2}{24} \left(\strut\SST + 2\TST\right)
        + \O(\dt^4)\) \\
      \hline
      QPQ & \2{\1\half T}{\1{}S}
        & \(T + S + \frac{\dt^2}{24} \left(\strut 2\SST + \TST\right)
        + \O(\dt^4)\) \\
      \hline
      \begin{tabular}{c}
	PQPQP \\
	\(\scriptstyle\alpha=\frac{1}{6}\) \\
	\cite{omelyan:2002,omelyan:2003,forcrand:2006}
      \end{tabular}
      & \3{\1{\rational16}S}{\1\half T}{\1{\rational23}S}
      & \(T + S + \frac{\dt^2}{72}\SST
        + \O(\dt^4)\) \\
      \hline
      \begin{tabular}{c}
	PQPQP \\
	\(\scriptstyle\alpha=\frac12\left(1-\frac1{\sqrt3}\right)\) \\
	\cite{omelyan:2002,omelyan:2003,forcrand:2006}
      \end{tabular}
      & \3{\1{\frac{3-\sqrt3}6}S}{\1\half T}{\1{\frac1{\sqrt3}}S}
      & \(T + S + \frac{\sqrt3-2}{24}\dt^2\TST
        + \O(\dt^4)\) \\
      \hline
      \begin{tabular}{c}
	Campostrini \\ PQPQPQP \\
	\cite{campostrini89a,creutz89a}
      \end{tabular}
      & \4{\1{\frac{\5{1}{2}{4}}{12}}T}%
          {\1{\frac{\5{1}{2}{4}}6}S}%
          {\1{\frac{\5{-1}{-2}{2}}{12}}T}%
          {\1{-\frac{\5{1}{2}{1}}3}S}
      & \(\begin{array}{c}
	  T + S \\[1ex]
	  + \rational{\dt^4}{34560} \left(
	  \begin{array}{c}
	    -(\5{40}{40}{48})\;\SSSST\\
	    +(\5{180}{240}{312})\;\STSST \\
	    +(\5{60}{80}{104})\;\STTST\\
	    +(\5{-20}{0}{8})\;\TSSST \\
	    +(\5{0}{20}{32})\;\TTSST \\
	    +(\5{0}{5}{8})\;\TTTST
	  \end{array}\right) \\[9ex]
          + \O(\dt^6)
        \end{array}\) \\
      \hline
      \begin{tabular}{c} Force \\ Gradient \\ PQPQP \end{tabular}
      & \3{\1{\rational16}S}%
	  {\1{\rational12}T}%
	  {e^{\frac{48\dt\;S - \dt^3\;\widehat{\SST}}{72}}}
      & \(\begin{array}{c}
	  T + S \\[1ex]
	  - \rational{\dt^4}{155520} \left(
	  \begin{array}{c}
	    41\;\SSSST \\
	    + 36\;\STSST \\
	    + 72\;\STTST \\
	    + 84\;\TSSST \\
	    + 126\;\TTSST \\
	    + 54\;\TTTST
	  \end{array}\right) + \O(\dt^6)
        \end{array}\) \\
      \noalign{\hrule height1.5pt}
    \end{tabular}
  \end{center}
  \vspace{-5mm}
  \caption{A selection of integrators with their exactly conserved shadow
    Hamiltonians.}
  \label{table:1}
\end{table}

\section{Computing the Force Gradient}

As shown in our previous proceedings \cite{Kennedy:2007cb,Clark:2008gh}
Poisson brackets for gauge theories may be written in terms of momenta \(p_i\)
and linear differential operators \(e_i\) that provide gauge-covariant
generalizations of the vector fields \(\partial/\partial q^i\); in particular
we have \(\SST = e^i(S)e_i(S)\).  The ``equations of motion'' for the
\(\widehat\SST\) vector field are \(\dot P = \widehat{\SST}P \defn-\G\) where
\(P\defn p_iT^i\), and we may use the relation \(e_i(U) = - T_iU\), where
\(T_i\) are the adjoint generators of the gauge group, to show that the Force
Gradient vector field is the ``second derivative'' \(\G = e^j(S)e_j
e_i(S)T^i\).

If we consider a generic pseudofermion action \(S = \phi^\dagger\M^{-1}(U)
\phi\) where \(\phi\) is a pseudofermion field, \(U\) the gauge field, and
\(\M\) any hermitian fermion kernel, then \[e_i(S) = -\phi^\dagger \M^{-1}
e_i(\M) \M^{-1}\phi = -X^{\dagger} e_i(\M)X\] where \(X \defn \M^{-1}\phi\).
The Force Gradient can be computed by applying the linear differential
operator \(\F\defn e^j(S)e_j\) to the above equation; by the Leibnitz rule
\[-\G = -\F\bigl(e_i(S)\bigr) = \F(X^\dagger)e_i(\M)X + X^\dagger\F\bigl(
e_i(\M)\bigr)X + X^\dagger e_i(\M)\F(X);\] defining \(Y\defn\F(X) = -\M^{-1}
\F(\M)X\), we obtain \[-\G = Y^\dagger e_i(\M) X + X^\dagger \F\bigl(e_i(\M)
\bigr) X + X^\dagger e_i(\M)Y.\] Note that the cost of computing the Force
Gradient in an HMC integrator is the inversion required to compute \(Y\) in
addition to the usual inversion needed to compute~\(X\).

\section{Results}

\begin{figure}
  \leavevmode%
  \epsfxsize=0.8\textwidth\centerline{\epsffile{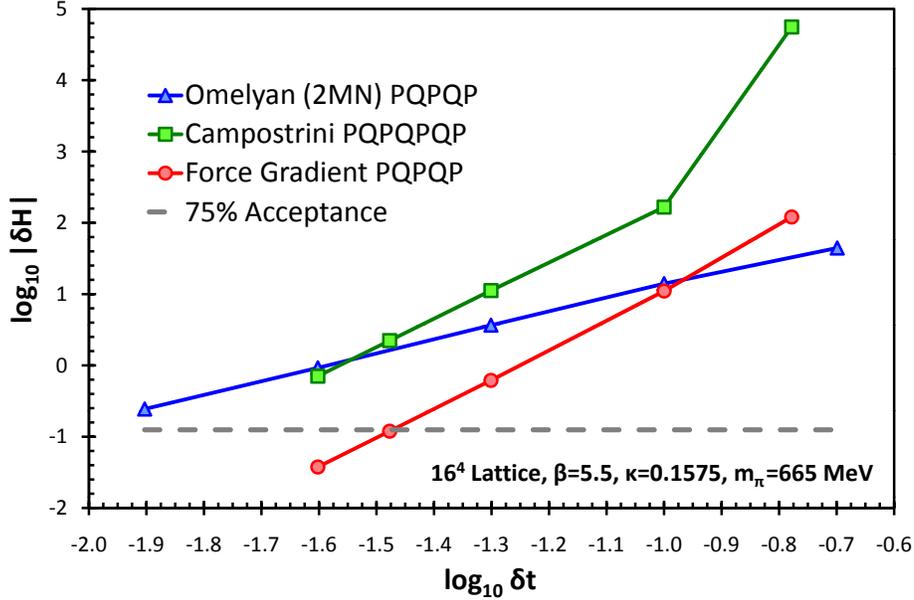}}
  \caption{Change \(\dH\) in the Hamiltonian over an entire trajectory as a
    function of the integration step size \(\dt\).  The dashed line indicates
    the value of \(\dH\) which corresponds to a 75\% HMC acceptance rate.}
  \label{fig:1}
\end{figure}

We have implemented this PQPQP Force Gradient integrator for lattice QCD with
dynamical Wilson fermions, and we present our initial results for a \(16^4\)
lattice at \(\beta=5.5\) and \(\kappa=0.1575\), which corresponds to a pion
mass of \(m_\pi = 665\)~MeV.  We used the usual even-odd preconditioning,
which does not introduce any significant complications into our formalism.

In Figure~\ref{fig:1} we show the change \(\dH\) in the Hamiltonian over an
entire trajectory as a function of the integration step size \(\dt\) on a
log-log plot.  To guide the eye we have also drawn a line to indicate the
value of \(\dH\) which corresponds to a 75\% HMC acceptance rate.  The slopes
of the lines correspond to the expected order of the integrators, and the
Force Gradient integrator is more than order of magnitude more accurate than
the Campostrini integrator at any step size, indicating that the coefficient
\(k\) discussed in the introduction is indeed much smaller.

\begin{figure}
\leavevmode%
  \epsfxsize=0.8\textwidth\centerline{\epsffile{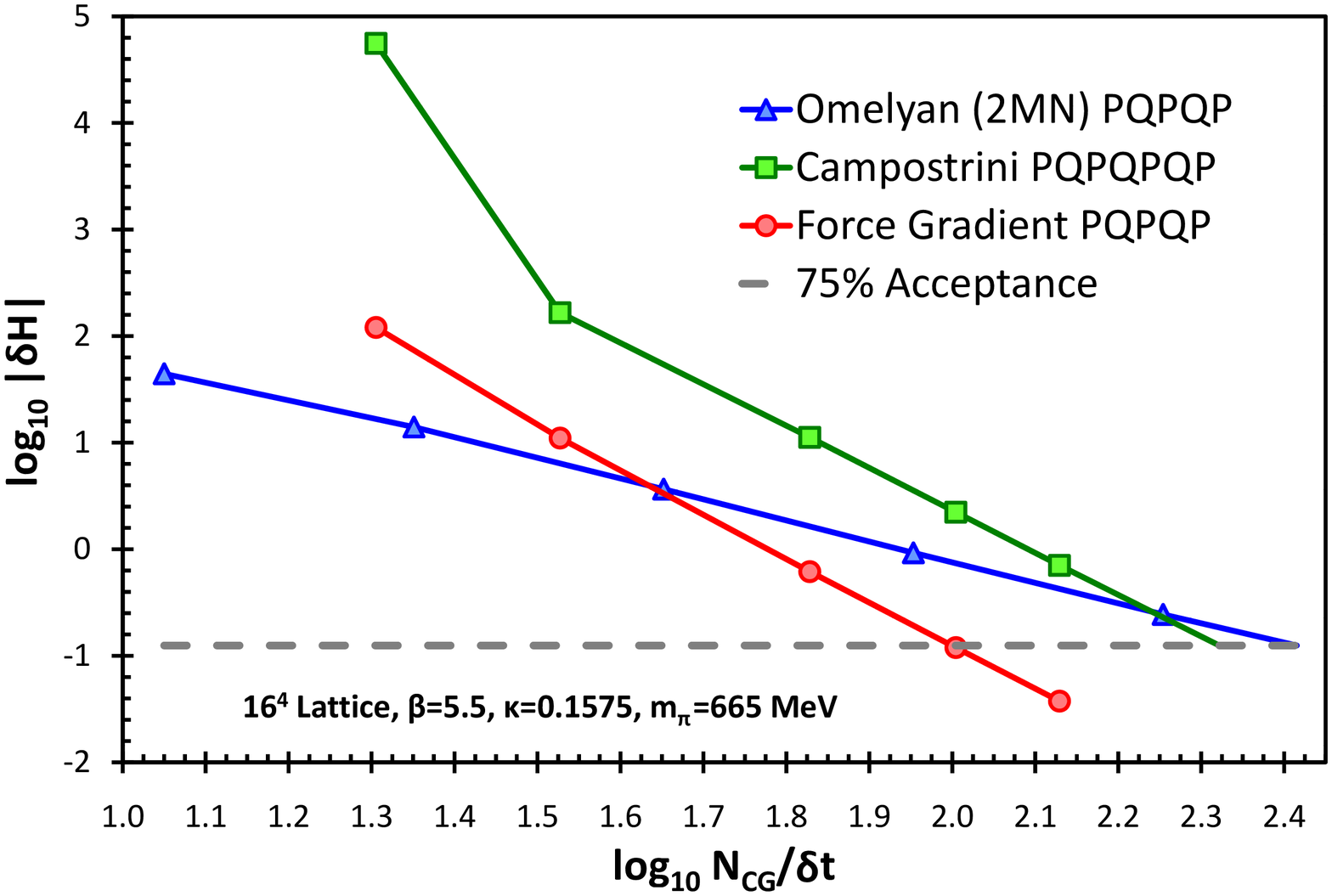}}
  \caption{Change \(\dH\) in the Hamiltonian over an entire trajectory as a
    function of an estimate of the computational cost.  The dashed line again
    indicates the value of \(\dH\) corresponding to a 75\% HMC acceptance
    rate.}
  \label{fig:2}
\end{figure}

In Figure~\ref{fig:2} we replot the same data as a function of an estimate of
the cost, namely the number of CG solutions divided by the step size.  The
Omelyan integrator requires two inversions of the Wilson--Dirac operator per
step, one\footnote{As initial and final ``half steps'' can be combined we
count each as half an inversion.} for each \(\hats\) (\(P\)) integration step,
whereas the Campostrini and Force Gradient integrators require three
inversions (one for each \(\hats\) and \(\widehat{\SST}\)).  From the
intercepts with the dashed line (75\% acceptance) we find that the Force
Gradient integrator is a factor of \(2.6\) cheaper than the Omelyan
integrator, which was up to now the preferred choice of integrator.

\section{Conclusions}

Our Force Gradient integrator is cheaper by more than a factor of two even for
small lattices with fairly heavy quarks, and the benefit increases as the
integration step size becomes smaller.  We expect that the integrators will be
improved by tuning using measured average values of Poisson brackets, as
described in~\cite{Clark:2008gh}.  We also note that our formalism for
computing Force Gradient integrators is compatible with all common actions,
smearing, and so forth.

\section*{Acknowledgements}

We would like to acknowledge support from FCT (SFRH/BPD/40998/2007),
NSF (PHY-0427646 and PHY-0835713), and the UK STFC.

\end{document}